\documentclass[runningheads]{llncs}
\usepackage[T1]{fontenc}

\usepackage{amssymb}

\usepackage{graphicx}
\usepackage{booktabs}   %
\usepackage{tabularx}
\usepackage{colortbl}
\usepackage{array}      %
\newcolumntype{L}[1]{>{\raggedright\arraybackslash}p{#1}}
\usepackage{hyperref}
\usepackage{xcolor}

\usepackage{cleveref}
\usepackage{todonotes}
\usepackage{enumitem}
\usepackage{microtype}

\usepackage{tikz}

\DeclareRobustCommand{\metfull}{\tikz[baseline=-0.6ex]\fill (0,0) circle (0.6ex);}
\DeclareRobustCommand{\metpart}{\tikz[baseline=-0.6ex]{\draw (0,0) circle (0.6ex);\fill (0,0) -- (90:0.6ex) arc (90:270:0.6ex) -- cycle;}}

\usepackage[section]{placeins}

\begin{document}
\title{On the Creation of Object-Centric Event Logs from Patient Journeys}
\titlerunning{On the Creation of Object-Centric Event Logs from Patient Journeys}
\subtitle{Research paper} %

\author{Saimir Bala\inst{1}%
\and
Neven Borak\inst{2}%
\thanks{First and second author contributed equally to this paper} \and
Maximilian König\inst{1}%
\and
Anjo~Seidel\inst{1}%
\and
Zeinab~Soleimani\inst{1}%
\and
Lorenzo Famiglini\inst{2}%
\and
Lev Petrov\inst{2}%
\and
Jonas~Witt\inst{2}%
\and
Mathias~Weske\inst{1}%
}
\authorrunning{S. Bala et al.}
\institute{Hasso Plattner Institute, University of Potsdam, Germany\\
\email{firstname.lastname@hpi.de}
\and
mama health, Germany, \url{https://www.mamahealth.com}\\
\email{firstname.lastname@mamahealth.io}\\
}
\maketitle              %
\begin{abstract}
Interactions between patients and health systems as well as conditions that individuals experience during a disease are represented in patient journeys. Therefore, patient journeys are important assets to investigate how care is being provided, and how health systems can be designed to better suit the needs of patients. While there is initial work on patient journeys from a process mining perspective with traditional case notions, object-centric process mining techniques have yet to be covered. This is unfortunate, since due to the rich structure of objects and their relationships, object-centric process mining approaches are particularly promising in this context. This paper addresses this gap by investigating object-centric event log (OCEL) generation from patient journeys. Based on a list of challenges and requirements, the paper presents an approach for the creation of OCELs from patient journey data. The approach is evaluated on a synthetic patient cohort provided by our industry partner.
\end{abstract}
\section{Introduction}
\label{sec:intro}

Process mining applied to healthcare can unlock unprecedented insights and help improve both the quality of care and the systems designed to support it~\cite{Arias2020,DBLP:journals/jbi/Munoz-GamaMFJSH22}. Typically, process mining has been applied to healthcare data from a healthcare provider's perspective. This perspective helps identify bottlenecks and inefficiencies in healthcare systems and improve their processes. However, it implicitly treats the patient as a ``customer'' of the process. This limits the scope of process improvement in three ways. First, it overlooks the patient's own knowledge, even though patients often possess extensive knowledge of their clinical histories. Second, due to data availability, it often restricts the analysis to individual information systems (e.g., those of a single hospital), whereas patient care spans multiple systems, clinicians, and treatments. Third, the insights from these analyses often aim to improve the healthcare provider's processes (e.g., those of a hospital). Such improvements may, of course, benefit patients. Nevertheless, the ultimate measure of the quality of care is what the patient experiences throughout their \emph{journey}~\cite{Oben2020}, which cannot be fully captured by analyzing only the data from information systems logs and must therefore be captured directly from patients themselves. 
However, such accounts are incomplete and may change, they are told in natural language, and they span a complex domain of conditions, treatments, and stakeholders. These three challenges make the data hard to model before analysis methods can be applied. Moreover, patient journeys involve several entities at once (e.g., patients, conditions, treatments). Object-centric event logs (OCELs)~\cite{berti2024ocel} are best suited to analyze all these perspectives at once.

In this paper, we identify challenges and requirements related to the creation of object-centric event logs from patient journeys. To address them, we propose a stepwise approach, starting from the data collection until the generation of an OCEL. Key features in our approach include the temporal and semantic resolution of the concepts extracted from the narrative of the patient. %
Our approach includes a mapping scheme to guide the creation of OCELs. This mapping scheme is based on general principles and can be applied to any patient journey to convert them to OCEL. We evaluate the approach on a synthetic cohort of patients.

The remainder of the paper is structured as follows. Section~\ref{sec:background} introduces patient journeys and their relation to processes, followed by a motivating example illustrating a patient journey. Section~\ref{sec:method} defines the challenges and requirements of analyzing a patient journey as a process, and presents an approach to address the creation of OCELs. Section~\ref{sec:evaluation} evaluates the approach. Section~\ref{sec:related} discusses related work. Section~\ref{sec:conclusion} summarizes the key findings.

\section{Background and Running Example}
\label{sec:background}

\subsection{Patient journeys}

From the perspective of a patient, care is a relationship with healthcare providers, such as clinics, doctors, or hospitals. It unfolds through a series of encounters with different doctors, locations, and treatments, which are connected by handovers~\cite{Vesinurm2024}.
A care process as it is planned is referred to as a \emph{pathway}. Thus, a \emph{clinical pathway} refers to the planned clinical interventions and therapies, along with the activities that enable and support them. When a clinical pathway is enacted in practice, we obtain a patient journey. A \emph{patient journey} is defined as what was done and what actually happened that affected the patient's medical condition, functioning, experience, and emotions~\cite{McCarthy2016,Oben2020}. 
At a higher level, a journey can be described as a sequence of clinical milestones, such as the symptom onset, the diagnosis, and the start of a treatment, and the transitions between them. Whether and when a patient reaches a milestone depends on factors, such as motivations, perceived barriers, and external triggers.
Modeling milestones and their factors shows where journeys stall on their way to a successful treatment.

Improving patients' experience requires considering their journeys as they actually occur. However, there is no central system that records the journey. Thus, data may be scattered across multiple systems (e.g., different hospitals, clinics, and doctor reports). Some steps may not be recorded at all (e.g., a session with a psychologist). This makes patients themselves the most knowledgeable stakeholders, and thus a key source of patient journey evidence. %
Collecting this evidence from patients is not trivial, as illustrated by the following example.

\subsection{Motivating Example}
\label{sec:example}

Myasthenia gravis (MG) is an autoimmune disease in which antibodies attack the neuromuscular junction, impairing the signal from nerve to muscle. Patients experience muscle weakness and rapid exhaustion with use, either throughout the body (generalized MG) or in specific muscle groups, such as those around the eyes or the bulbar muscles that control swallowing, speaking, chewing, and breathing. Treatments either relieve symptoms by inhibiting acetylcholinesterase or suppress the immune attack. Below, we introduce a running example of a fictional patient's journey that is affected by MG.

\subsubsection{A patient's journey.}
Lena teaches school in Hamburg. Most days she is fine until about one o'clock; in the last two lessons her voice thins and turns nasal, and word went around the school that she had been drinking at lunchtime.

It began with a tiredness that sleep did not fix, a year or more before anything specific happened. The first thing she could point at was her jaw tiring halfway through a meal. An ear, nose, and throat (ENT) doctor found her vocal cords normal, put it down to tension, and sent her to speech therapy; after six sessions, the speech therapist was the first to say that something did not add up.
By early 2023, she was choking on water and had lost seven kilos; one bad episode took her to the emergency department, which sent her home the same night. She was referred to neurology and waited four months. The test for the most common antibody was negative and nerve stimulation was normal, but her blood tests showed that the Hashimoto's thyroiditis diagnosed in 2019 was undertreated. Hypothyroidism accounted for everything, her levothyroxine was increased, and her thyroid levels came back normal. Nothing else did. Five months later, a specialist she paid to see privately tested a second antibody, muscle-specific kinase (MuSK): positive, in November 2023, nearly two years after she first mentioned any of it. ``I cried, and then I was furious for about six months.'' She kept the discharge letter, but not the documentation from the earlier visits that failed to diagnose her.

Pyridostigmine gave her cramps and was abandoned after three months. Steroids worked, but she gained eleven kilos and lost her sleep and the shape of her face. Azathioprine followed. In February 2025, an on-call doctor treated a urinary infection with an antibiotic, and within two days she could not lift her head off the pillow: nine days in hospital with immunoglobulin infusions. After two rounds of rituximab, she is back at work four days a week, better than at any point since 2022.

\section{Methodology}
\label{sec:method}

\subsection{Challenges \& Requirements}
\label{sec:requirements}

The running example illustrates three key challenges. We call each clinical event that a patient reports, such as a visit, a diagnosis, or a treatment, a \emph{fact}. The term \emph{event} refers to an event in an event log (e.g., OCEL).

\noindent\textbf{C1: Open world.} There is uncertainty in the data. A journey is known only through what is reported, and what is reported is incomplete and may change. Lena kept the discharge letter but not the documentation of the earlier visits, her first symptoms date from ``a year or more before'', and later reports may correct earlier ones. A fact that is not reported may still have happened, and only an explicit statement tells that something did not.

\noindent\textbf{C2: Natural language.} Journeys are told in the patient's own words. The same concept is described in different words and at different levels of detail, e.g., ``her voice thins and turns nasal'' for a speech symptom, or ``steroids'' for a specific drug. One must decide which concept a statement refers to.

\noindent\textbf{C3: Complex domain.} A journey spans several conditions, treatments and stakeholders, documented in heterogeneous sources. Lena's journey runs through MG, Hashimoto's thyroiditis and a urinary infection, involves an ENT doctor, a speech therapist, the emergency department, neurology and a private specialist, and is documented in her narrative and a discharge letter. Its facts influence each other. Patients, clinicians, and researchers look at the same journey from different perspectives.

From the challenges C1--C3, we derive six requirements. R1, R2, and R4 address \emph{open world} (C1), R3 addresses \emph{natural language} (C2), and R5 and R6 address \emph{complex domain} (C3). C3 calls for multi-perspective analysis, e.g., with OCELs.

\noindent\textbf{R1: Traceable evidence (C1).}
What is known about a journey changes, as new evidence refines, corrects, or contradicts earlier reports. Hence, every fact must remain \emph{traceable} to the evidence it was obtained from and to when it was reported or revised.

\noindent\textbf{R2: Uncertain time (C1).}
Patients do not tell their journey in chronological order. They jump back and forth in time, give ranges (``by early 2023'') or relative times (``five months later''), report durations without dates, or specify no time at all. Lena's early episode, from the first tiredness to the speech therapy, is known only in the order in which it is told. An OCEL, however, requires a timestamp for each event. Hence, narrative order, temporal order, and timestamps must be distinguished, and facts whose time cannot be resolved must not be lost.

\noindent\textbf{R3: Normalized terminology (C2).}
Patient-reported concepts must be \emph{normalized} against clinical terminologies while keeping the patient's own words.

\noindent\textbf{R4: Non-occurrence (C1).}
What did not happen can be as relevant as what did. A patient stating that she never had a thymectomy conveys different information than a thymectomy not being mentioned. A planned visit differs from one that took place. Hence, negated and intended facts must be represented and distinguished from occurred ones.

\noindent\textbf{R5: Multiple perspectives (C3).}
Fine-grained facts are needed to preserve the individual history. Higher-level abstractions are needed to compare journeys across patients and to identify common patterns, bottlenecks, and delays. Hence, journeys must be analyzable from several perspectives, such as conditions and treatments, and at several levels of abstraction.

\noindent\textbf{R6: Related facts and factors (C3).}
Patients relate facts to each other. The ENT diagnosis led to speech therapy, and the crisis followed the antibiotic. Progression through the journey may also depend on factors, such as barriers, motivators, and triggers (e.g., specialist gatekeeping or fear of side effects). OCEL~2.0 relates events only implicitly, through the objects they share~\cite{berti2024ocel}. Hence, relations and factors must be represented, and patient-reported associations must be distinguished from established clinical causality.

\subsection{Approach for Creating OCELs from Patient Journeys}
\label{sec:approach}

\begin{figure}[t]
    \centering
    \includegraphics[width=.85\linewidth]{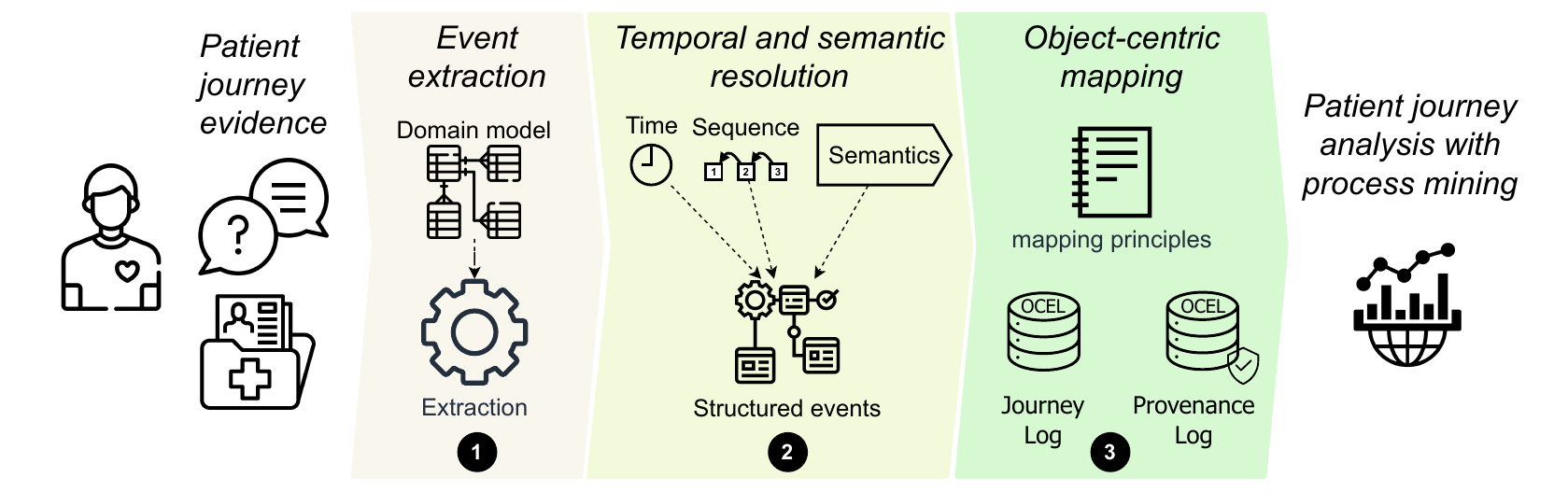}
    \caption{A stepwise approach for constructing OCELs from patient journeys.}
    \label{fig:approach}
\end{figure}

Fig.~\ref{fig:approach} illustrates the steps of our approach. We assume that patient-journey evidence is collected from interviews, questionnaires, and clinical records, and possibly by AI agents. In step~{1}, facts are extracted from this evidence. A domain model defines which event types, relations, and clinical concepts are extracted. Each fact cites the evidence it was extracted from and keeps its history of revisions (R1), and is marked as occurred, negated, or intended (R4). In step~{2}, the extracted facts are resolved temporally and semantically: reported times become date ranges and temporal relations (R2), and patient wording is matched to clinical terms (R3). In step~{3}, the structured clinical events (i.e., facts) are mapped to two OCELs, a journey and a provenance log, according to a set of mapping principles. On top of the journey log, milestones and transitions provide higher-level abstractions. The resulting log can then be analyzed with object-centric process mining techniques.

\subsection{Event Extraction and Domain Modeling}

\begin{figure}
    \centering
    \includegraphics[width=.85\linewidth]{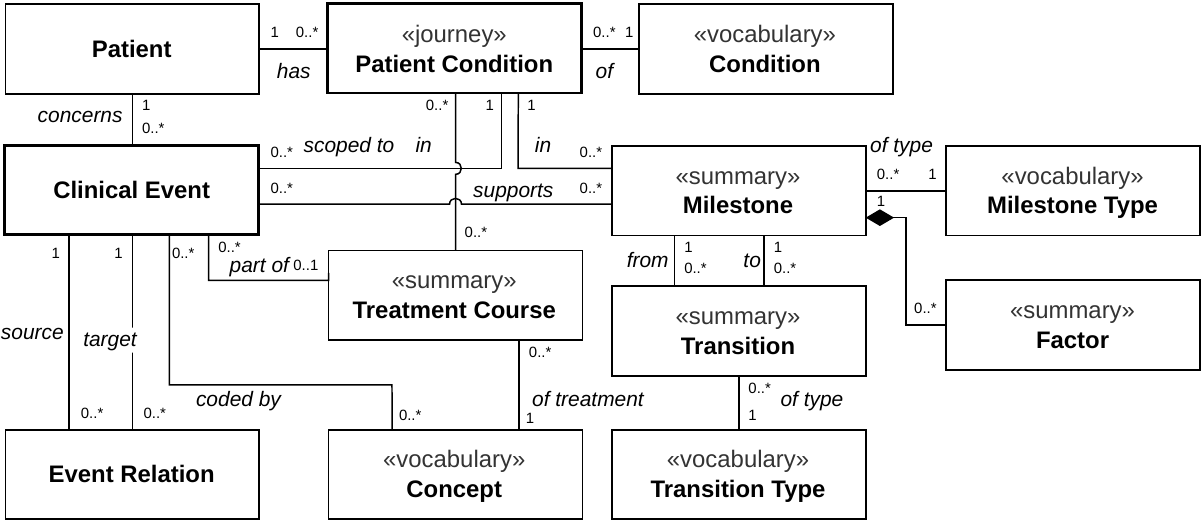}
    \caption{Domain model of a patient journey. \guillemotleft vocabulary\guillemotright: shared codes, \guillemotleft summary\guillemotright: derived from the facts.}
    \label{fig:domain-data-model}
\end{figure}

In order to extract structured information from the patient journey evidence, we need to specify \emph{what} is extracted. This is done through a domain model. The domain model defines the event types, inter-event relations, standard terms, and other relevant entities that need to be extracted from the evidence. \Cref{fig:domain-data-model} shows a UML domain model, based on the one used by our industry partner.  

We explain the model using the running example.
Lena is a \textsf{Patient} and has reported several \textsf{Conditions}, i.e., MG, Hashimoto's thyroiditis, and a urinary infection. Each combination of patient and condition, called a \textsf{Patient Condition}, is treated as a journey.
Everything Lena reports, such as the ENT visit or the speech therapy, is a \textsf{Clinical Event}, which we call a fact. A clinical event belongs to Lena and to any number of her journeys (e.g., a visit at which several conditions are discussed), and it is coded by \textsf{concepts} from a shared vocabulary. Clinical events can be linked to each other by \textsf{Event Relations}. Each relation has a source, a target, and a kind, which is either causal or temporal. For instance, the antibiotic \emph{causes} the myasthenic crisis, whereas the speech therapy \emph{follows} the ENT visit. From the clinical events, a summary of each journey is derived. Lena's pyridostigmine treatments form one \textsf{Treatment Course}. The positive MuSK test supports the \textsf{Milestone} of reaching the diagnosis, whose Milestone Type is diagnosis. It was preceded by a misdiagnosis milestone, so a \textsf{Transition} links the two. A transition expresses temporal order, not causation. \textsf{Factors} record what drove or hindered reaching a milestone. For example, doctor unfamiliarity with MG and the rarity of the disease hindered timely diagnosis and treatment.

\subsection{Temporal and Semantic Resolution}
\label{sec:resolution}

Narratives move back and forth in time, and patients describe clinical concepts in their own words. The goal of this step is to resolve facts with respect to time, sequence, and semantics.

\noindent\emph{Time.} Reporting order is not occurrence order. The time at which a fact occurred is therefore represented as a date range with a precision, unless the exact date is reported. For instance, ``by early 2023'' becomes a range from January to March 2023 with month precision.
Reported durations are kept, so that the end timestamp of a fact can be derived from its start (e.g., pyridostigmine was ``abandoned after three months'' (R2)), if the start becomes known.

\noindent\emph{Sequence.} Where date ranges are insufficient to order facts, a large language model (LLM) extracts temporal relations from the narrative sequence. This is particularly useful to resolve which facts were concurrent (e.g., to define treatment combinations (R2)). In the running
example, Lena reports that azathioprine ``followed'' the steroids. Only a temporal relation tells whether it replaced them or was added to them.

\noindent\emph{Semantics.} Clinical terms such as symptom and treatment names are resolved via fuzzy matching against a vocabulary database, while the patient's own words are kept (R3). For instance, ``her voice thins and turns nasal'' is matched to a speech symptom. Any relevant shared vocabulary can be used for this purpose%
. In this work, we use a custom vocabulary actively used by our industry partner. %

\subsection{Object-Centric Mapping}
\label{sec:mapping}

This step maps the structured facts to OCEL~2.0.

\noindent\textbf{M1: Facts are objects, their occurrences are events (R2, R4).}
Each fact becomes an object whose type is the event type of the domain model, e.g., \emph{Treatment} or \emph{HcpVisit}. An OCEL event is created only if the fact occurred and its time is known: a point fact yields one event, an interval fact a start event and, if its end is known or follows from a reported duration, an end event. The timestamp is the start of the reported date range; the range and its precision are kept as event attributes. In the running example, Lena's pyridostigmine treatment yields a \emph{StartTreatment} event in November 2023 and an \emph{EndTreatment} event three months later. Her visit to the ENT doctor has no date, and her planned outpatient reviews have not taken place. Hence, they remain in the log as objects without events. 
Facts that report a change, e.g., a stopped treatment or an improved symptom, become events on the object they change. Stopping the antibiotic becomes a \emph{StopTreatment} event on the treatment course, and a change of disease status becomes a new value of the \emph{disease status} attribute of the journey.
Milestones are mapped like facts. Each milestone becomes an object, and a reached milestone with a known time also yields an event, e.g., \emph{ReachDiagnosis}. A milestone that is not reached (explicitly denied) remains an object without an event, whereas an unknown milestone (not mentioned) is not in the log. The factors of a milestone become its attributes.

\noindent\textbf{M2: Relations between facts become relations between objects (R6).} 
As OCEL~2.0 relates events only through objects, a relation between two facts becomes an O2O relation between their objects, with the relation type as qualifier. For example, the antibiotic treatment \emph{causes} the myasthenic crisis.
The qualifier names the kind of relation. Whether a relation was reported by the patient or established clinically is not kept, since O2O relations carry no attributes. Keeping it would require an object for each relation.
Transitions between milestones become O2O relations in the same way.

\noindent\textbf{M3: Perspectives become object types (R5).}
A perspective is a kind of entity whose behavior we want to follow over time, similar to the case notion in traditional process mining. Each perspective becomes an object type, and each of its entities an object. An event is related to the fact that it records and, through that fact, to the objects of each perspective to which it belongs. Hence, the same events can be analyzed from different perspectives. In the running example, patients, journeys, and treatment courses are perspectives: the start of Lena's azathioprine belongs to her, to her MG journey, and to her azathioprine course. 

\noindent\textbf{M4: Terms are normalized, wording is kept (R3).}
Each fact keeps the normalized term and the patient’s own words as attributes. Catch-all codes, such as ``other treatment'', are not used as concepts, as they would merge unrelated facts. Instead, the concept is derived from the normalized wording or omitted. 

\noindent\textbf{M5: Log occurred events, keep reporting events in a separate log (R1).}
An OCEL attribute value has a single timestamp. We therefore create two logs. The \emph{journey log} uses the time at which facts occurred. The \emph{provenance log} uses the time at which they were reported. It holds every fact, with or without an event in the journey log, and records the messages, questionnaire answers, and uploaded documents the facts come from, and when each fact was asserted or revised. In both logs, each fact refers to the evidence it was extracted from.

\subsection{Patient Journey Analysis}\label{sec:analysis}
The OCEL can be analyzed from each perspective (M3). In our setting, these are the patient, the journey, and the treatment course. 
A case-centric log offers only one perspective, e.g., the patient. A higher level of abstraction does not remove this limit. A treatment event remains an event of the patient, even if its course runs next to another one, and a diagnosis of MG and one of asthma remain the same activity. Milestones and transitions provide one level of abstraction at which journeys can be compared across patients. Process mining on milestones shows the most common pathways, as well as bottlenecks, delays, and drop-offs. %

\section{Evaluation}\label{sec:evaluation}

We applied the approach to the journeys of a cohort of synthetic patients. The data come from synthetically generated evidence from mama health, which mimics real-world patient journeys that cannot be published due to privacy protection of personally identifiable health data. The data comprise 280 synthetic patients with MG (4 further patients without a journey summary were left out). They report 13{,}684 facts and 2{,}267 milestones. From them, we created two logs. The journey log holds each fact as an object, except 72 changes of disease status, which are events on their journey (M1). Of the 13{,}684 facts, 7{,}346 occurred at a known time and yield 8{,}364 events, e.g., the start and end of a treatment. The other 6{,}338 facts (46.3\%) are negated, intended, or without a usable time, and remain objects without events. Together with the milestones (2{,}000 events), 280 patients, 800 journeys, and 747 treatment courses, the journey log holds 17{,}706 objects and 10{,}364 events of 30 activities. The provenance log holds all 13{,}684 facts again, each asserted or revised at its reporting time, together with the 12{,}254 messages, 2{,}795 questionnaire answers, and 446 documents they come from. It has 29{,}179 events and 30{,}553 objects. A replication package can be found on GitHub~\footnote{\url{https://github.com/mama-neven/OCEL-patient-journeys-replication-package}}.

\subsubsection{Requirements.}\label{sec:eval-requirements}

\begin{table}[tbp]
\centering
\caption{Requirements versus principles; \metfull: met by construction, \metpart: met in part}\label{tab:requirements}
\footnotesize
\newcolumntype{L}{>{\raggedright\arraybackslash}X}
\newcolumntype{P}[1]{>{\raggedright\arraybackslash}p{#1}}
\begin{tabularx}{\textwidth}{@{}P{2.4cm}P{2.9cm}P{0.9cm}L@{}}
\toprule
Requirement & In OCEL & Met & Evidence \\
\midrule
R1 Traceable evidence & evidence references, provenance log & \metfull\ M5 & All 13{,}684 facts refer to their evidence.\\
R2 Uncertain time & date ranges as attributes, undated facts as objects & \metpart\ M1 & 956 facts (7.0\%) without a resolved time remain objects. \\
R3 Normalized terminology & terms and words as attributes & \metfull\ M4 & Every term and wording is kept.\\
R4 Non- occurrence & objects without events & \metfull\ M1 & 5{,}377 facts (39.3\%) are negated or intended. \\
R5 Multiple perspectives & perspectives as object types & \metfull\ M3 & All 7{,}564 facts that the source links to a condition relate to their journey. \\
R6 Related facts and factors & O2O relations, factors as attributes & \metpart\ M2 & 7{,}184 relations are kept, but not whether they are patient-reported or clinical. \\
\bottomrule
\end{tabularx}
\end{table}

The logs meet a requirement when the information described in it is kept after the mapping, in the journey log or, for R1, in the provenance log, so that it can be analyzed. For each requirement, Table~\ref{tab:requirements} states how the log keeps this information and whether it is met. Some requirements are met by construction if a mapping principle guarantees them for every fact, so our check only confirms that the implementation applies the principle. We check each requirement on the logs and compare the counts with those of the source data, as reported by the mapping engine. 

Two requirements are met only in part. R2 is limited by OCEL itself. An event has a single timestamp, so a date known only to the month becomes the first day of that month. As a result, 31.0\% of the directly-follows pairs of a patient share their timestamp, and PM4Py orders them arbitrarily. 
R6 does not keep the basis of a relation (M2). The other four requirements are met by construction.

\begin{figure}[htbp]
\centering
\includegraphics[width=.7\textwidth]{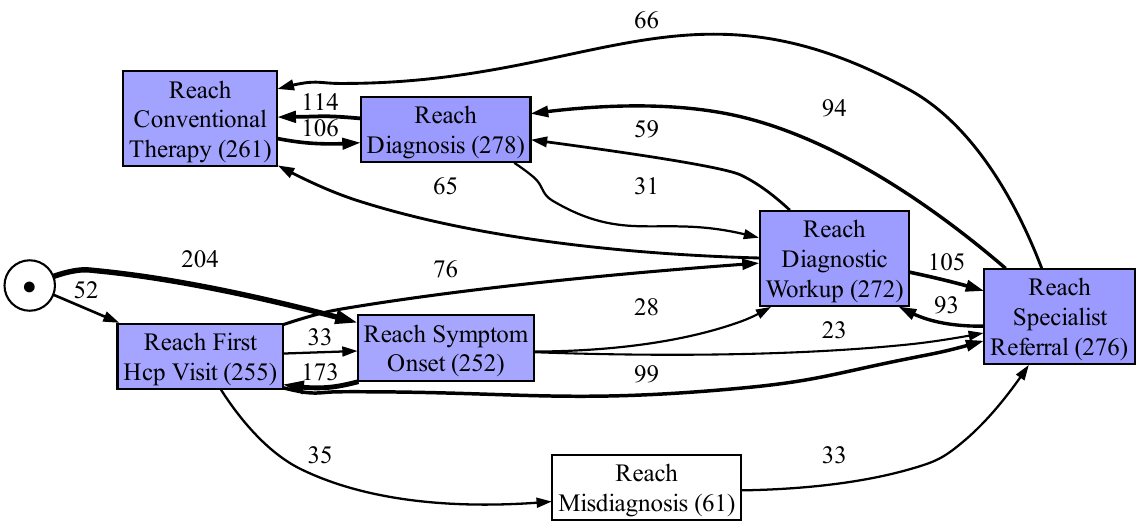}
\caption{Directly-follows graph of the milestones up to conventional therapy, with the patient as case (PM4Py). Edges with fewer than 23 pairs are hidden.}
\label{fig:journey-dfg}
\end{figure}

\subsubsection{Patient Journey.}\label{sec:eval-case-centric}

We used PM4Py to flatten the journey log on the patient. The result is a case-centric log with 280 cases and all 10{,}364 events, each exactly once.  Restricted to the milestones, it matches the log of the journey summaries that our industry partner analyzed before. Its 2{,}000 events of 15 milestone types form 204 variants. 
PM4Py discovers the directly-follows graph in Fig.~\ref{fig:journey-dfg}. The usual journey measures follow from the same log. The median time from symptom onset to diagnosis is 12.7 months (IQR 7.0--18.3, $n=251$). It is 19.6 months for patients with a misdiagnosis ($n=57$) and 10.6 months for the others ($n=194$). 
Our industry partner confirmed that these results match its own analysis of the journey summaries. 

\subsubsection{Object-Centric Process Mining.}\label{sec:eval-object-centric}

\begin{figure}[bp]
\centering
\includegraphics[width=.7\textwidth]{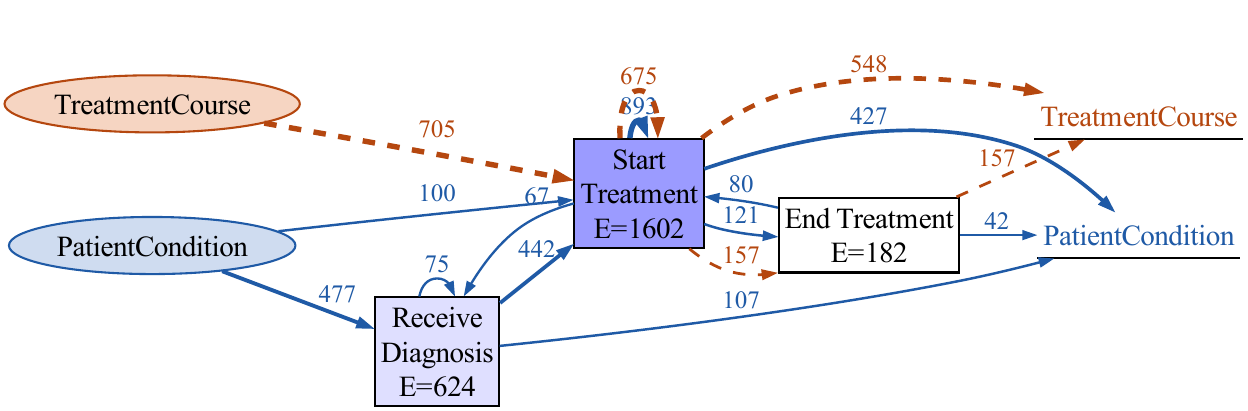}
\caption{OC-DFG of the journey log (PM4Py), restricted to diagnosis, treatment start, and treatment end. Solid: journeys, dashed: treatment courses.}
\label{fig:journey-course-ocdfg}
\end{figure}

A patient's journeys and treatment courses are objects of their own.
In the journey log, 211 of the 280 patients reported journeys of comorbidities next to MG.
With the patient as case, a diagnosis of MG and one of asthma are the same activity, \texttt{ReceiveDiagnosis}.
With the journey as case, the link to the overarching patient is lost, and so are the 38.5\% of fact events not associated with a journey.
The object-centric log keeps both. We used PM4Py to discover the OC-DFG in Fig.~\ref{fig:journey-course-ocdfg} from the journey log as it is. Journeys and treatment courses share the treatment activities, but not their paths.
477 journeys start with a diagnosis, while a course starts with a treatment. 
After its end, a course stops: its path never returns from \texttt{EndTreatment} to \texttt{StartTreatment} (0 pairs). 
The path of a journey returns 80 times, in 67 journeys. Each time, the treatment that starts belongs to another course of the same journey (80 of 80). 

The figure thus shows treatment switches, although no activity is named so. With the patient as case, the DFG has the same edge 114 times. It cannot tell a switch from a restart or from a treatment of another condition.

\subsubsection{Discussion.}\label{sec:eval-discussion}
Some information is still lost. In particular, 53 occurred facts state a duration but have no end event. PM4Py drops 6{,}771 objects without events and the 4{,}712 O2O relations that end in one of them. An analysis of what did not happen must read the log itself. Providers are kept as attributes of visits. A provider perspective would require provider identities, which the data do not contain. 
\emph{Threats to validity.} The data are synthetic and may share assumptions with the domain model. The cohort is not held out in a strict sense. All patients were selected for MG, and the milestones and treatment courses are defined for MG only. We did not assess faithfulness against an independent ground truth.

\section{Related Work}
\label{sec:related}

Munoz-Gama et al.~\cite{DBLP:journals/jbi/Munoz-GamaMFJSH22} describe the characteristics of healthcare processes and identify challenges for analyzing them. Arias et al.~\cite{Arias2020} and Shafei et al.~\cite{Shafei2024} treat patient journey analysis as a customer journey mapping problem and analyze the patient journey with process mining. Both build their event logs from information systems and use a single case notion. In contrast, we start from patient-reported evidence and create OCELs. 

Object-centric process mining has been applied to MIMIC-IV, from a database that complies with the OMOP common data model~\cite{DBLP:journals/jbi/ParkLC24}, including conformance checking~\cite{Tripathi2024}. Bala et al.~\cite{BalaSW26} use LLMs to generate event logs and show that incorrect medical terms limit the result. These works start from structured clinical databases. There, facts (i.e., events in the domain) neither lack a time nor fail to happen, and they are not revised later.

For the creation of event logs in general, Seidel et al.~\cite{Seidel2026} outline challenges, and Fahland
et al.~\cite{DBLP:journals/corr/abs-2410-14495} propose a core model and discuss its design space. Other approaches extract logs from databases supporting SAP ERP~\cite{DBLP:journals/jiis/BertiPRA23}, with a goal-driven
method~\cite{DBLP:journals/dke/Wei0WH25}, or by extending the PM$^2$ methodology~\cite{DBLP:conf/bpmds/MiriKZ025}. Jans and Soffer~\cite{DBLP:conf/bpm/JansS17} show that creating an event log involves many decisions, and Swevels et al.~\cite{Swevels2026} propose process-mining-aware domain models. 
All of them start from data recorded by information systems. %
Our mapping principles start from patient-reported evidence, and keep facts that have no timestamp or did not happen.

\section{Conclusion}
\label{sec:conclusion}
This paper addressed the creation of OCELs from patient journeys. From three challenges, we derived six requirements and proposed a stepwise approach. Its mapping principles keep each fact as an object and create an event only when a fact occurred at a known time. We applied the approach to a synthetic cohort of 280 patients with MG. The resulting logs conform to OCEL~2.0 and keep the 46.3\% of facts that have no event. They also separate treatment courses that a case-centric log merges. R2 and R6 are met only in part. R2 is limited by single timestamps, and R6 by plain O2O relations.
Future work will apply the approach to real patient journeys and other conditions, and evaluate the extraction and resolution steps against a ground truth.

\bibliographystyle{splncs04}
\bibliography{refs}
\end{document}